\documentclass[aps,floatfix,prd,showpacs]{revtex4}
\usepackage{graphicx}
\usepackage{amsmath}
\usepackage{dcolumn}
\usepackage{bm}

\begin{document}

\def\be{\begin{equation}}
\def\ee{\end{equation}}

\title{Interpretation of scalar and axial mesons in LHCb from a historical perspective}

\author{F. E. Close}
\affiliation{
Rudolf Peierls Centre for Theoretical Physics, University of Oxford, 
Oxford, OX1 3NP, UK.}

\author{A.Kirk}
\affiliation{Culham Centre for Fusion Energy, Abingdon, Oxfordshire, U.K.
}


\date{\today}

\begin{abstract}
LHCb measurements of 
$B_{d,s} \to J/\psi + X$ 
are shown to be consistent with
historical data  on scalar and axial mesons below 2 GeV.
This is in contrast to some recent interpretations of these data. 
Further tests of our hypotheses in other
$B_{u,d,s} \to J/\psi + X$
decay modes are suggested.

\end{abstract}

\maketitle

\section{Introduction}

Recently the LHCb collaboration has released a series of papers 
\cite{lhcbbskk}
\cite{lhcbb4pi} 
\cite{lhcbbspipi}
\cite{lhcbbdkk} 
\cite{lhcbbdpipif980}
\cite{lhcbbdpipinof980}
on  $B_d$ or $B_s$ decays to $J/\psi + X$ where $X$ is a scalar or axial vector meson. This mechanism gives a potentially clean flavour filtering. 
Production in $B_s \to J/\psi + X(s\bar{s})$ occurs at leading order and the 
$B_d \to J/\psi +X(d\bar{d})$ is the dominant entree to the accompanying meson $X$, though Cabibbo suppressed. 
Thus $B_s$ gives an amplitude proportional to the $s\bar{s}$ content of $X$, and the $B_d$ decay gives amplitude proportional to its $d\bar{d}$ content. There is no direct way to probe the $u\bar{u}$ content by recoil against $J/\psi$, 
but decays of the charged $B^{-} \to J/\psi X^-$ give information about the coupling to strange (Cabibbo leading) 
or isovector states (Cabibbo suppressed).

 LHCb has investigated the flavour content of $X$ for both scalar mesons and for the $f_1(1285)$ by measurement of 
$B_{d,s} \to J/\psi + X$. 
The conclusions, prima facie, differ from those drawn from experiments at CERN by both WA102 (central production), 
GAMS and Crystal Barrel (proton antiproton annihilation at low momentum). 
Our interest was triggered by how similar some of the features observed in the LHCb data are to what was found in the 1990s.
Our strategy is to adopt the conclusions from the aforementioned experiments and 
see if they are consistent with the LHCb data. We then identify specific further tests that may shed light on any outstanding issues.

The basic conclusions and their relevance to the LHCb results, are as follows:


1. CERN hadronic experiments (such as central production, $p\bar{p}$ annihilation at LEAR, etc) indicated the presence of three scalars in the 1.3 to 1.8 GeV mass range. Interference between these, and the different production mechanisms in the various processes and channels, explained why peaks appeared with different mass or width in various channels in different experiments.

2. LHCb has observed peaks in this mass region in the decays of $B_{d,s} \to J/\psi 2\pi; J/\psi 4\pi$. 
Here too there may be indications that the masses and widths are channel dependent. 

3. We have taken the masses, widths and the same logic as in the old CERN results. 
(We have even used the same computer programmes as were used in WA102). 
We made no adjustments to the input. 
We then compared the output to the published data of LHCb where available. Despite the fact we have no access to the acceptance corrected data, the
description of the raw mass spectra are remarkably consistent. It would be interesting to see how robust this description would be if compared with the
acceptance corrected data.

4. Further, if we use the model interpretation of the three scalars $f_0(1370);f_0(1500); f_0(1710)$ as mixtures of a standard nonet $n\bar{n};s\bar{s}$ and a scalar glueball, $G$, which historically 
fitted existing data on hadronic and $J/\psi$ decays, \cite{cqz} the relative normalisation of some signals can be assessed. 

First we use the $f_1(1285)$, as seen in $B_{d,s} \to J/\psi 4\pi$. The LHCb results are found to be consistent with the flavor mixture deduced historically, where, approximately: $f_1(1285) \sim 0.9n\bar{n} - 0.4 s\bar{s}$ \cite{ckf1}. This allows predictions to be made for other channels, 
specifically the production of scalar mesons above 1 GeV in $B_{d,s} \to J/\psi 4\pi$. 
Where it is possible to do so from published data, everything is consistent: the conclusion is that 
LHCb is consistent with world data on these scalar and axial mesons.

5. We then use these results to look at $B_{d,s} \to J/\psi 2\pi$. 
If we assume presence of $f_0(980)$ consistent with the upper limit obtained by LHCb, then the solution is consistent with historical results.
In particular, there is no need for more radical conclusions, 
reported in \cite{cern}.

6. We propose an interpretation of LHCb data consistent with significant isospin violation in the vicinity of $f_0(980)$ and the $K\bar{K}$ threshold.
This phenomenon has been noted historically in hadron data\cite{ads,as,krs,wa102scalar,ck00}. 
A test of this hypothesis is proposed for LHCb in the channel $B_s \to J/\psi a_0(980)$.

\section{$B_{d,s} \to J/\psi 4 \pi$}

Even without any detailed analysis, the similarity between LHCb and WA102 spectra on $4\pi$ is evident.
First compare the $4\pi$ spectra from LHCb (Figs 2 of Refs.~\cite{lhcbb4pi}) with that from WA102\cite{wa1024pi}. A clear signal of $f_1(1285)$
is visible in both cases, as well as strength above 1.3 GeV. In the case of the $B_d$ decay the spectrum also shows clear structure centred around 1.45GeV, noticeable for having a sharp rise on the low mass side and a gradual fall on the high mass side. 
This looks similar  to what  WA102 observed in the 1990s  and described as being due to the interference between two scalar 
mesons: $f_0(1370)$ and $f_0(1500)$ Ref.~\cite{wa914pi}. First we compare the data on $f_1(1285)$ and establish their quantitative consistency.

\subsection{$B_{d,s} \to J/\psi f_1(1285)$}

If the flavor mixing basis is defined by
\begin{eqnarray}
f_1(1285) =cos\theta \frac{1}{\sqrt{2}}|d\bar d + u\bar{u}\rangle + sin\theta | s \bar s \rangle 
 \label{eq:f1mix}
\end{eqnarray}
\noindent then the ratio of branching ratios
\begin{eqnarray}
\frac{B_d \to J/\psi f_1(1285)}{B_s \to J/\psi f_1(1285)} = \frac{ cot^2\theta}{50} 
\label{eq:f1mixtheta}
\end{eqnarray}
\noindent where we have assumed that the phase space for $B_d$ and $B_s$ decays are the same and that their lifetimes are also. (In practice this is correct to within 2\%). We have approximated the Cabibbo suppression $tan^2\theta_c \sim 1/25$.

A previous analysis by  
Close and Kirk, performed using the data from WA102\cite{ckf1}, showed that the flavour content of the $f_1(1285)$ was  
$f_1(1285) \sim n\bar{n} - 0.4s\bar{s}$. This solution is consistent with a flavor mixing angle of 22$^o$. For pedagogic illustration, take this angle $\theta$ to be half of 45degrees, and hence $cot^2\theta = \frac{1+\sqrt{2}}{1-\sqrt{2}} = 5.7$.
Thus Eq.~(\ref{eq:f1mixtheta}) would predict the ratio of rates $\sim 11.4\%$ to be compared with LHCb data: $11.6 \pm 3.1\%$.
Thus we conclude that the LHCb data agree with the historical solution.

\subsection{$B_{d,s} \to J/\psi f_0(1370/1500)$}

The similarity in structure of the $4\pi$ spectrum in LHCb with that of WA102, and the quantitative agreement with the
$f_1(1285)$ signal, inspires a comparison of the full spectrum, and of the enhancement around 1450MeV with the interference solution
advocated by WA102.

\section{Fits to the mass spectra}
\subsection{Using the parameters from WA102 to fit the $B_s  \rightarrow J/\psi  4\pi $ data from LHCb}
A fit has been performed to the the $4\pi$ mass spectrum using the parameters obtained from the $\pi^+\pi^-\pi^+\pi^-$ mass spectrum from WA102 \cite{wa1024pi} namely
\begin{tabbing}
XXXXXXXXXXXXXX \= XfX1(1285):X \= mX =X 1285 MeVXXX   \= $\Gamma$ = 20 MeV  \kill
\> $f_1(1285):$ \> m = 1285 MeV   \> $\Gamma$ = 20 MeV  \\
\> $X(1450):$   \> m = 1445 MeV   \> $\Gamma$ = 100 MeV  \\
\> $f_2(1950):$ \> m = 1910 MeV   \> $\Gamma$ = 450 MeV 
\end{tabbing}

\begin{figure}[ht]
\centerline{\includegraphics[width=12cm]{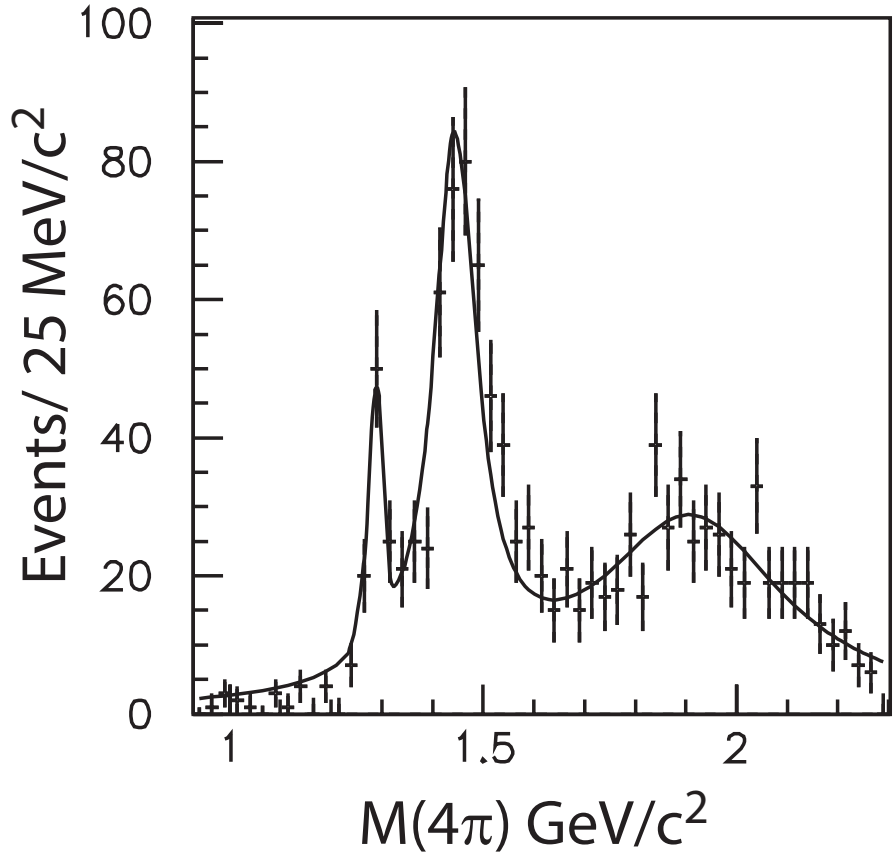}}
\caption{Fit to the LHCb $4\pi$ mass spectrum obtained from $B_s  \rightarrow J/\psi  4\pi $ with the parameters obtained from the fit to the WA102 Mass spectrum including an $X(1450)$.}
\label{fi:x1450}
\end{figure}

\noindent The relative amplitude of the states is the only free parameter. The resulting comparison with LHCb data is shown in Figure \ref{fi:x1450}. 
It should be noted that we have no absolute normalisation and have illustrated a fit to the total published
mass spectrum. In particular, we have assumed that the region around 1450 MeV is dominated by J=0.
LHCb indicate the presence of some helicity 1 in this region, this would change the normalisation but not the shape of our fits.
However, we urge that careful re-analysis of this region be made in view of the fact that there is no known state
of the required mass and width with J>0 decaying to 4pi in the 1450 MeV region.
As stated earlier, our aim is not to produce a perfect fit to the published data but rather to suggest a possible method that the LHCb collaboration 
could apply to their data.

\begin{figure}[ht]
\centerline{\includegraphics[width=12cm]{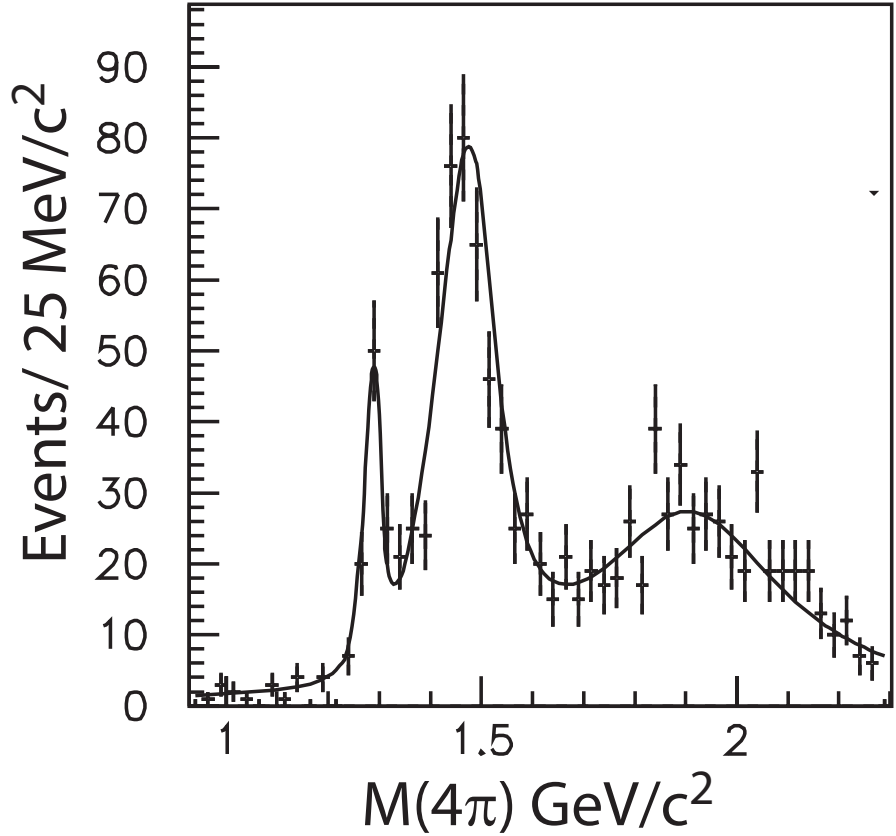}}
\caption{Fit to the LHCb $4\pi$ mass spectrum obtained from $B_s  \rightarrow J/\psi  4\pi $ with the parameters obtained from the fit to the WA102. The mass spectrum includes 
interference between the $f_0(1370)$ and $f_0(1500)$.}
\label{fi:bs4piint}
\end{figure}

Similar to what has been observed in LHCb, experiment WA102 and its predecessor WA91 \cite{wa914pi}  found that although a peak in the 1500 MeV mass 
region was observed in several decay channels it appeared to have a mass and width that was channel specific. 
 At a similar time in the 1990s other experiments were also observing peaks in this region. 
For example, the GAMS collaboration observed a $G(1590)$ decaying to $\eta\eta$ and $\eta\eta^\prime$ \cite{GAMS}. 
Whilst the Crystal Barrel experiment observed the $f_0(1500)$ in several final states in $p\bar{p}$ annihilations \cite{CRYSTALBARREL},  
it was realised that these discrepancies could be solved by assuming that two scalar states in this region interfered to produce the observed peaks.  
The different production rates and decay rates to different channels led to distinct interference structures which explained all the observations.  
The two states have become known as the $f_0(1370)$ and $f_0(1500)$.  
Our fit to the $4\pi$ LHCb mass spectrum from the reaction $B_s  \rightarrow J/\psi  4\pi $ allowing for the interference between these two states 
as well as the $f_1(1285)$ and $f_2(1950)$
is shown in Figure \ref{fi:bs4piint}. 
In this fit we have constrained the masses and widths of the $f_0(1370)$ and $f_0(1500)$  to lie within $1 \sigma$ of the values obtained in WA102. The resulting masses and width from the fit are
\begin{tabbing}
XXXXXXXXXXXXXX \= XfX1(1285):X \= mX =X 1285 MeVXXX   \= $\Gamma$ = 20 MeV  \kill
\> $f_0(1370)$: \> m = 1340 MeV \> $\Gamma$ = 200 MeV \\
\> $f_0(1500)$: \> m = 1490 MeV \> $\Gamma$ = 135 MeV
\end{tabbing}
Given the constraints on the parameters used in the fit
and uncertainties in background contributions, this parametrisation gives a reasonable description of the data.

\subsection{Predicting the $4\pi$ mass spectrum in $B_d  \rightarrow J/\psi  4\pi $ data from LHCb}

We have used the $s \bar s$  and $n \bar n$ coupling of the  $f_1(1285)$ and the number of events observed 
in $B_d  \rightarrow J/\psi  4\pi $ and $B_s  \rightarrow J/\psi  4\pi $ 
to calculate a normalisation for the two channels. 
Using (a) this normalisation, (b) the number of $f_0(1370)$ and $f_0(1500)$ obtained from the fit to the  $B_s  \rightarrow J/\psi  4\pi $ 
and (c) the previously determined coupling of the $f_0(1370)$ and $f_0(1500)$  to $s \bar s$  and $n \bar n$ 
determined from the WA102 data (see below) \cite{closekirkscalars}, we can predict the number of events that should be observed in $B_d  \rightarrow J/\psi  4\pi$. The only free parameter is the
relative mixing angle between the two states, which may be production mechanism dependent, and the contribution from the $f_2(1950)$.

In order to predict $B_d  \rightarrow J/\psi  4\pi $ we need a model for the flavour content of scalar mesons. Our purpose is to test the consistency of LHCb data with models abstracted from historical data. In the global perception of the latter we distinguish those below and above 1 GeV. The template for the low-lying scalars is based on Jaffe\cite{jaffe}

\begin{eqnarray}
[ud][\bar u\bar d] &&  f_0(500) \nonumber \\
{}[ud][\bar d \bar s],\ [ud][\bar s \bar d],\ [us][\bar u \bar d],\ [ds][\bar d \bar u]  &&  \kappa \nonumber \\
\frac{1}{\sqrt{2}}([su] \bar s\bar u] + [sd][\bar s\bar d]) && f_0(980) \nonumber \\
{}[su][\bar s \bar d],\  \frac{1}{\sqrt{2}}([su][\bar s \bar u] - [sd][\bar s \bar d]),\  [sd][\bar s \bar u]  &&  a_0(980).
\label{eq:diq}
\end{eqnarray}

Those above 1 GeV are consistent with a $n\bar{n}$ and $s\bar{s}$ nonet mixed with a scalar glueball. This is motivated by lattice QCD, and remains consistent with  data accumulated for over a decade. For convenience, we summarise the
resulting trio of isoscalar $0^{++}$ states as follows

\begin{eqnarray}
f_0(1370) = +0.6|G\rangle - 0.1|s\bar s \rangle - 0.8 | n \bar n \rangle \nonumber \\
f_0(1500) = -0.7|G\rangle + 0.4|s\bar s \rangle - 0.6 | n \bar n \rangle \nonumber \\
f_0(1710) = +0.4|G\rangle + 0.9|s\bar s \rangle + 0.15 | n \bar n \rangle 
\label{eq:scalarGns}
\end{eqnarray}

The absolute values should not be taken too seriously, but the relative division into large, medium and small, is robust as are the relative phases: 
all constructive in 1700; the destructive phase between $s\bar s$ and $n \bar n$ in 1500, and the constructive versus destructive phases between $G$ and $n \bar n$ in the three cases.

\begin{figure}[ht]
\centerline{\includegraphics[width=12cm]{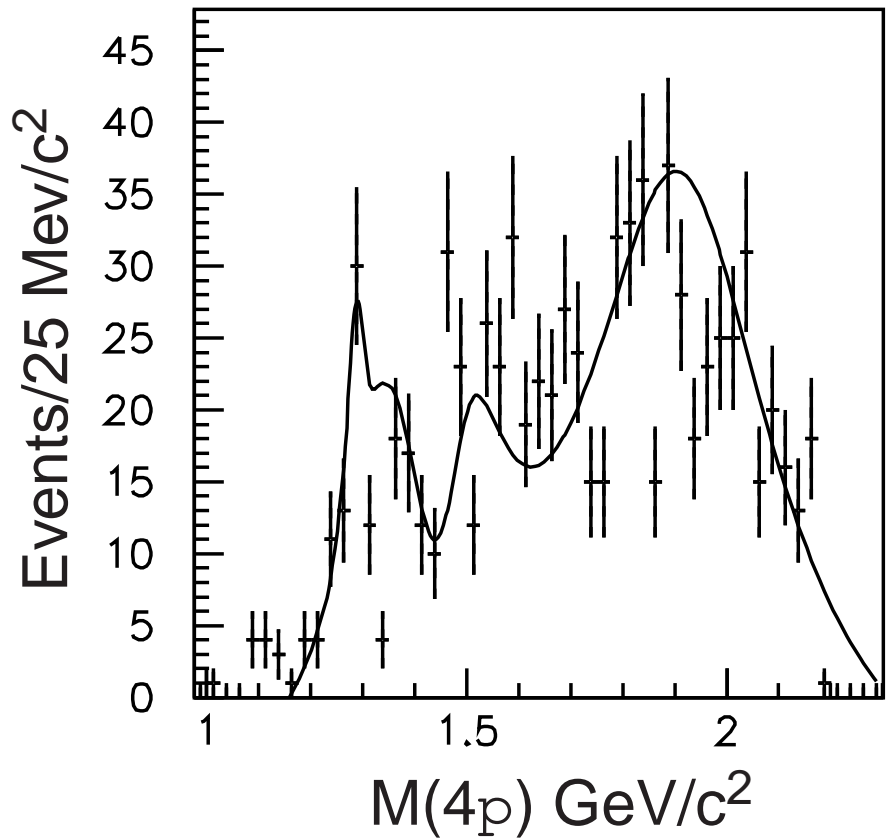}}
\caption{Fit to the LHCb $4\pi$ mass spectrum obtained from $B_d  \rightarrow J/\psi  4\pi $ with the parameters obtained from the 
$B_s$ decay mode 
and the previously determined coupling of the $f_0(1370)$ and $f_0(1500)$ to $s \bar s$  and $n \bar n$ .}
\label{fi:bd4piint}
\end{figure}

The resultant prediction for $B_d  \rightarrow J/\psi  4\pi $  is shown in Figure \ref{fi:bd4piint}, which 
is a fair representation of the empirical mass spectrum.

We now describe the quantitative comparison of these data with the model.


\section{Scalar meson model above 1.2 GeV and LHCb data}

Here we show the model expectations that led to the curves in Fig. \ref{fi:bd4piint}.
The resulting fit is shown in Figure \ref{fi:bskkfit}a. 
The scalars are produced in proportion to their $d\bar{d}$ and $s\bar{s}$ content, analogous to the analysis of the $f_1(1285)$ case.
We can thus make statements about their relative strengths in both $B_d$ and $B_s$ decays to $J/\psi f_0$.
Ignoring phase space (the error in this is probably less than that in the intrinsic flavour strengths anyway!), the relative production rates are as follows:

\begin{eqnarray}
B_s \to J/\psi f_0(1370): f_0(1500):f_0(1710) = 1:16:81 
\label{eq:scalarrates}
\end{eqnarray}
\noindent and from $B_d$ decays:

\begin{eqnarray}
B_d \to J/\psi f_0(1370): f_0(1500):f_0(1710) = 32:18:1
\label{eq:scalarrates0}
\end{eqnarray}
\noindent The relative rates from $B_d$ (per unit of $d\bar{d}$ in the amplitude) are Cabibbo suppressed by a factor of 
about 25 relative to those of $B_s$ (per unit $s\bar{s}$).

These rates will be perturbed, however, by the presence of $G$. 
Thus, an extra coupling strength in $B_s \to f_0(1370)$ may be expected, for example. 
Although there will be a suppression factor $x$ for $B_s \to 0.6G \to 0.6xs\bar{s}$, this overall may compare with $B_s \to 0.1s\bar{s}$. Inclusion of this gluon contribution gives relative contributions:

\begin{eqnarray}
B_s \to J/\psi [ f_0(1370): f_0(1500):f_0(1710)] = (1+6x)^2:(4+7x)^2:(9+4x)^2
\label{eq:gluemix}
\end{eqnarray}
\noindent We ignore the glue admixture in the $B_d$ decays as the direct couplings to $d\bar{d}$ are either 
large or comparable to the glue content, and so the latter is at most a perturbation, to the accuracy of our schematic model.

The relative rates in $B_{d,s} \to J/\psi 4\pi$ are given by Eqs.~(\ref{eq:scalarrates}) and (\ref{eq:scalarrates0}) after 
the respective branching ratios $f_0 \to 4\pi$ are taken into account. 
Approximately the BR of $f_0(1500) \sim 50\%$ whereas that of $f_0(1370) \sim 100\%$ \cite{wa102f0br}. 
Thus the relative orders of magnitude that we would expect in this simple model (ignoring glue) are

\begin{eqnarray}
B_s \to J/\psi 4\pi[f_0(1370): f_0(1500):f_0(1710)] = 1:8:0
\label{eq:scalarrates4pi}
\end{eqnarray}
\noindent Empirically, the curves in Figure \ref{fi:bs4piint}. 
give ratio $1: 6.4 \pm 0.6 \pm 1.0:0$

For $B_d$ decays the model then implies:

\begin{eqnarray}
B_d \to J/\psi 4\pi[ f_0(1370): f_0(1500):f_0(1710)] = 32:9:0
\label{eq:scalarrates04pi}
\end{eqnarray}
\noindent and these were imposed to generate the curves.

\subsection{Possible evidence for the $f_0(1710)$ in $B_s$ decays}


Having shown that the data from LHCb on $B_{s,d} \rightarrow J/\psi  4\pi $ are consistent with the presence of the $f_0(1370)$ and $f_0(1500)$, we now turn our attention to the other member of this scalar triplet, namely the 
$f_0(1710)$. 
This state decays prominently to $KK$ and $\eta\eta$ , and was not observed to decay to $4\pi$ in WA102.  The 
relative decay rates were \cite{wa102f0br}:\\
\begin{tabbing}
xxxxxxxxxxxxxxx \= f0(1222)XXX   \=   pi piXX \= : XXXXXK  K \= : XXXXXetaeta : \= XXXXXetaetaprime \= : XXXXX4pi  \kill
\> \> $ \pi \pi$ \> : $K \overline K$ \> : $\eta\eta$ \> : $\eta\eta^\prime$ \> : $4\pi $\\ 
\>$f_0(1710)$: \> $1$ \> : $5.0 \pm 0.7$ \> : $2.4 \pm 0.6$ \> : $<\;0.18\;(90\;\%\;\; CL)$ \> :$ <\;0.54\;(90\;\%\;\; CL) $\\
\end{tabbing}
Our fits have assumed that the branching ratio $f_0(1710) \rightarrow 4\pi = 0$.

The canonical model implies that this state has a big production rate in $B_{d,s} \to J/\psi s\bar{s}$ but small decay width to pions, as observed. It implies a significant signal in $J/\psi K\bar{K}$, also as seen empirically. As this channel also has considerable interest for a complete understanding of the production of the $f_0(980)$, we initially focus on the mass region above 1.2 GeV and return to the 1 GeV  region later.

\begin{figure}[ht]
\centerline{\includegraphics[width=12cm]{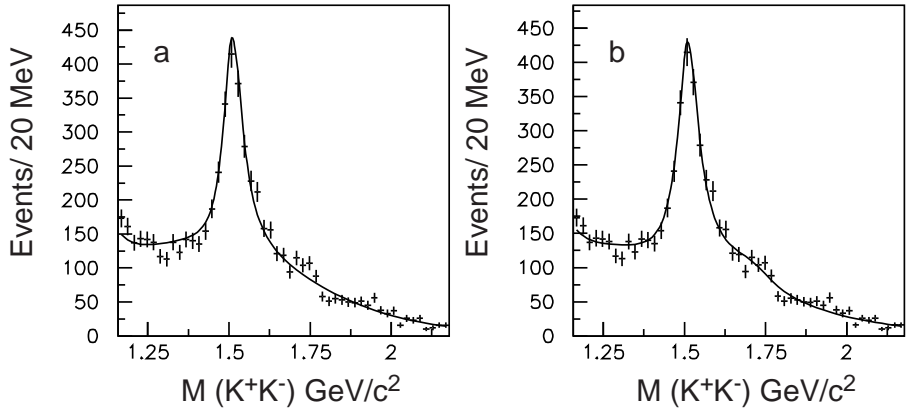}}
\caption{Fit to the LHCb $K^+K^-$ mass spectrum obtained from $B_s  \rightarrow J/\psi  K^+K^- $ a) without and b) with the parameters of the $f_0(1710)$ obtained by WA102}
\label{fi:bskkfit}
\end{figure}

We first turn our attention to the reaction 
$B_s  \rightarrow J/\psi  K^+K^- $, which is dominated by the $\phi(1020)$ and $f_2(1525)$ \cite{lhcbbskk}. 
The resulting fit is shown in Figure \ref{fi:bskkfit}a. 
The fit can be improved further by including a contribution from 
the $f_0(1710)$ of the PDG~\cite{pdg}, with
mass and width fixed to the PDG values namely: $m$ = 1722 MeV $\Gamma$ = 135MeV. 
The resulting fit is shown in Figure \ref{fi:bskkfit}b. The $\chi^2/NDF$ has decreased from 80/65 to 69/63 or the probability has improved from 0.09 to 0.28. 

\begin{figure}[ht]
\centerline{\includegraphics[width=12cm]{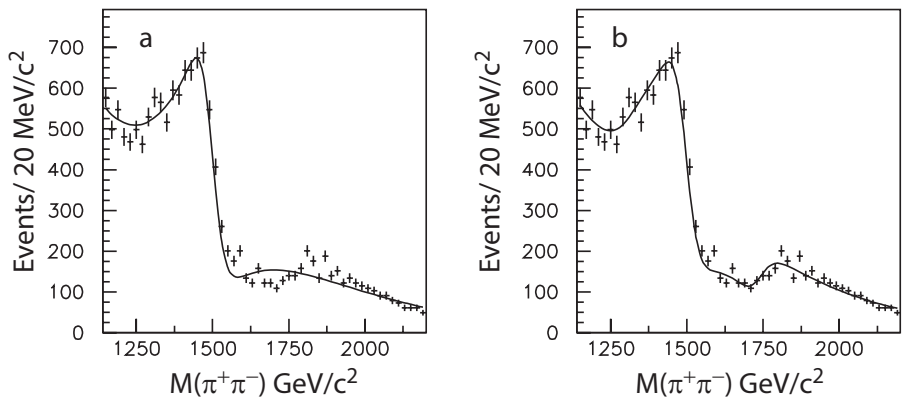}}
\caption{Fit to the LHCb $\pi^+\pi^-$ mass spectrum obtained from $B_s  \rightarrow J/\psi  \pi^+\pi^- $ a) without and b) with the inclusion of the 
 $f_0(1710)$}
\label{fi:bspipi}
\end{figure}

Fits have also been performed to the $\pi^+\pi^-$  mass spectrum above 1.2 GeV
observed in the decays $B_s  \rightarrow J/\psi  \pi^+\pi^- $ \cite{lhcbbspipi}
 using one of the parametrisation used to fit the S-wave $\pi^+\pi^-$ spectrum in WA102 \cite{wa102pipi}. 
 In this fit, the mass and width of the $f_0(1370)$ and $f_0(1500)$ and $f_0(1710)$ have been fixed to the WA102 values, 
which are consistent with those of the PDG~\cite{pdg}.
Their relative amplitude and phase have been left free, 
and interference between these states was allowed. 
The results of the fit to the region above 1.2 GeV is  shown in 
Figure \ref{fi:bspipi}a and b without and with the inclusion of the $f_0(1710)$ respectively. 
An improved fit to the mass spectrum is obtained, and the presence of the 
$f_0(1710)$ is clear. Thus we see evidence that this state is present in the $B_s$ decays in both the $K^+K^-$ and $\pi^+\pi^-$ channels.

Having shown that LHCb data above 1.2 GeV are consistent with the canonical picture, we suggest that LHCb,
who have access to the acceptance corrected data sets, reanalyse their data incorporating the scalar resonances we have discussed.
We now turn to the scalar mesons below 1 GeV where the WA102 fit included a low mass $\pi^+\pi^-$  
enhancement (or $\sigma$(500)), and $f_0(980)$.



Having established an interpretation of the 4$\pi$ data with three scalar mesons above 1 GeV, 
we can now examine the implications of this picture for $B_{d,s} \to J\psi 2 \pi$. This will also have structure below 1 GeV, 
linked to the $\sigma(500)$ and $f_0/a_0(980)$ in the scalar meson sector.


\section{Light scalars and Isospin Breaking}

 In the specific case of $B_{d,s} \to J/\psi f_0(980)$ there is an important issue about the role of the nearby $K\bar K$ threshold, 
which, inter alia, leads to effects that are in effect isospin violating. 
This has not been taken into account in existing analyses of the LHCb data and, as we shall now show, are critical in their interpretation. 
 
 First we briefly review the LHCb results on the $f_0(980)$.
The original paper \cite{lhcbbdpipif980} includes $f_0(980)$ in the fit. 
The subsequent paper \cite{lhcbbdpipinof980}, with approximately twice as many statistics, concludes that the $f_0(980)$ is not required statistically. 
Nonetheless, by eye one it is possible to see a discrepancy between the data around 1 GeV and the fit which indicate the 
possible presence of the $f_0(980)$ (for example, Figs. 12 and 13 of Ref.~\cite{lhcbbdpipinof980}). 
Based on an upper limit for the production of the $f_0(980$, Reference \cite{SZ} has investigated the implications for models of scalar mesons. 
We generalise their analysis to take account of possible isospin violation in the $f_0 - a_0 -K\bar{K}$ threshold system, which was determined historically.    
Remarkably, this could be the most significant effect in this region. 
We now review this, discuss its implications, and propose a further test for LHCb.

The $K\bar{K}$ threshold plays an essential role in the existence and properties of the $f_0(980)$ and $a_0(980)$. 
Isospin symmetry is broken by the $u-d$ quark mass difference. 
The mass gaps between these scalar mesons and the $K^+K^-$ relative to $K^0\bar{K^0}$ thresholds differ significantly. 
The S-wave couplings of the scalar mesons to the $K\bar{K}$ threshold thus lead to asymmetric couplings to the $u\bar u$ and $d\bar d$ flavors. 
As remarked above, the potential for significant flavor asymmetry, manifested as isospin violation, in these states has long been recognised.\cite{ads,as,krs}. Direct empirical evidence for a significant flavour distortion 
was found in the central production of scalar mesons by WA102\cite{wa102scalar}, where an isoscalar coupling to the $a_0$ was found to occur with an intensity of order 10\% of the canonical isovector. 

Whether these states are $K\bar{K}$ molecules, or instead the $K\bar{K}$ threshold merely drives these effects, is still to be resolved. 
In either case, the coupling of $f_0(980)$, for example, to $u\bar u$ and $d\bar d$ will differ in strength. 
In Ref.~\cite{ck00} the relative effective couplings to $K\bar{K}$ threshold was defined

\begin{eqnarray}
f_0(980) = cos \theta |K^+K^-\rangle + sin \theta | K^0\bar{K^0} \rangle \nonumber \\
a_0(980) =  sin \theta |K^+K^-\rangle - cos \theta | K^0\bar{K^0} \rangle 
\label{eq:kk}
\end{eqnarray}

\noindent and the mixing angle empirically determined to be $\theta = 30 \pm 3^o$. 
For pedagogic purposes and analytic illustration, we use $\theta=\pi/6$, which is consistent with our result. Thus we may
describe the relative amplitudes for coupling of these scalar mesons to light flavors as follows:

\begin{eqnarray}
f_0(980) = \frac{1}{2}|d\bar d \rangle + \frac{\sqrt{3}}{2} | u \bar u \rangle \nonumber \\
a_0(980) = \frac{\sqrt{3}}{2} | d \bar d \rangle - \frac{1}{2}|u\bar u \rangle 
\label{eq:ispin}
\end{eqnarray}

The production of mesons in $D,B$ and $D_s,B_s$ decays has traditionally been recognised as a means to assess the light flavour content of said mesons. 
LHCb has recently reported data on $B_d \to J/\psi +f_0(d \bar d)$ and $B_s \to J/\psi + f_0(s \bar s)$, where $f_0$ refers to either $f_0(980)$ or $\sigma(500)$, and the flavors quoted in parentheses are theoretically the dominant entree to the appearance of these states.

The presence of $\sigma$ in  $B_d \to J/\psi +\sigma(d \bar d)$ and its absence in $B_s \to J/\psi + \sigma(s \bar s)$ is consistent with this state having strong affinity for $u\bar{u}$ and/or $d\bar{d}$ without need for $s\bar s$. This is all as expected, given its established affinity for $\pi^+\pi^-$, and its relative distance from the $K\bar{K}$ threshold. 
The $f_0(980)$ however, while seen in $B_s \to J/\psi + f_0(s \bar s)$ was not yet visible in $B_d \to J/\psi +f_0(d \bar d)$. 
The analysis in Ref.~\cite{SZ} showed this to be consistent with the $f_0(980)$ and $\sigma(500)$ being mixtures of $q\bar q$ flavour states 
as in Eq. (\ref{eq:mix}). 

\begin{eqnarray}
|f_0\rangle &=& \phantom{-}\cos\varphi |s\bar s\rangle + \sin\varphi |n\bar n \rangle \nonumber \\
|\sigma \rangle &=& -\sin\varphi |s\bar s\rangle + \cos\varphi |n\bar n \rangle\nonumber \\
|n\bar n\rangle &=& \frac{1}{\sqrt{2}}(|u\bar u\rangle + |d \bar d\rangle),
\label{eq:mix}
\end{eqnarray}
In this case the amplitude for $\bar B_d\to J/\psi f_0$ is proportional to $\sin\varphi/\sqrt{2}$, while $\bar B_d \to J/\psi \sigma$ is proportional to $\cos\varphi/\sqrt{2}$.


Alternatively, if the $f_0$ and $\sigma$ are diquonium states as in Eq. (\ref{eq:diq}), then -- it was argued --  the $f_0$ should be produced with relative strength $1/\sqrt{2}$, while the $\sigma$ is produced with relative strength 1. Thus Ref. \cite{SZ} obtained the predictions

\be
\frac{BF(\bar B_d \to J/\psi \, f_0(980))}{BF(\bar B_d \to J/\psi \, \sigma)}  = \tan^2\varphi \cdot \frac{\Phi(f_0)}{\Phi(\sigma)}
\ee
for scalars with $q\bar q$ structure, and

\be
\frac{BF(\bar B_d \to J/\psi \, f_0(980))}{BF(\bar B_d \to J/\psi \, \sigma)}  = \frac{1}{2} \cdot \frac{\Phi(f_0)}{\Phi(\sigma)}
\label{eq:tq}
\ee
for scalars with $qq\bar q \bar q$ structure. Here $\Phi$ is a phase space factor.

The measured value for this ratio was based on a fit to the Dalitz plot for the reaction $\bar B_d \to J/\psi \pi^+\pi^-$\cite{lhcbbdpipinof980}. This fit found no evidence for the $f_0(980)$ and hence a small value of the mixing angle was reported:

\be
\tan^2 \varphi < 0.098 \ \rm{at}\ 90\%\ {\rm C.L.}
\ee
This value is 8 sigma removed from the diquonium prediction of Eq. (\ref{eq:tq}), hence the LHCb collaboration concluded that the diquonium picture of the light scalar nonet is strongly disfavoured.

These conclusions, however, all made the assumption that the $f_0(980)$ couples to $d\bar d$ and $u\bar u$ with equal strength. As argued above, and in Refs. \cite{ads,as,krs,ck00}, this is simplistic. The data imply that the coupling of $f_0(980)$ to $d\bar d$ is suppressed, and the empirical flavour basis of Eq.~(\ref{eq:ispin}) already qualitatively leads to such an expectation.

As an illustration, we now repeat the analysis of Ref. \cite{SZ} but with the light flavor basis of Eq.~(\ref{eq:ispin}). Present data are consistent with this. 
An immediate consequence, of course, is that a significant production of $a_0(980)$ should now arise. Obtaining evidence of this state will be a defining test for this hypothesis.

With the states defined as in Eq.~(\ref{eq:kk}), the relative couplings to $B_d$ are as follows (Ref.~\cite{SZ} corresponds to $\theta = \pi/4$).

\begin{eqnarray}
\langle d \bar d |f_0 \rangle & = & sin \theta \nonumber \\
\langle d \bar d |a_0 \rangle & = & cos  \theta\nonumber \\
 \langle d \bar d | \sigma \rangle & = & 1,
\label{eq:damps}
\end{eqnarray}

The analogous amplitudes for $B_s$ are

\begin{eqnarray}
\langle s \bar s |f_0 \rangle & = & sin \theta  + cos \theta\nonumber \\
\langle s \bar s |a_0 \rangle & = & sin \theta - cos  \theta\nonumber \\
 \langle s \bar s | \sigma \rangle & = & 0,
\label{eq:samps}
\end{eqnarray}

The generalisations of the amplitudes defined in Ref.~\cite{SZ} become

\begin{eqnarray}
  r^{s0f_0}_{sf_0} \equiv \frac{\Gamma(B \to f_0)}{\Gamma(B_s \to f_0)} = \frac{sin^2 \theta}{(cos \theta + sin \theta)^2} \nonumber \\
 r^{0f_0}_{0\sigma} \equiv \frac{\Gamma(B \to f_0)}{\Gamma(B \to \sigma)} =  sin^2 \theta \nonumber \\
  r^{s\sigma}_{s f_0} \equiv \frac{\Gamma(B_s \to \sigma)}{\Gamma(B_s \to f_0)} = 0 \nonumber\\
  r^{sf_0}_{0\sigma} \equiv \frac{\Gamma(B_s \to f_0)}{\Gamma(B_d \to \sigma)} = (cos \theta + sin \theta)^2
  \label{eq:amps}
\end{eqnarray}

The analysis of Ref.~\cite{SZ} has assumed that  $\theta = \pi/4$; however, the relative production of $f_0$ and $a_0$ by gluons in central production favours $\theta \sim \pi/6$.  With this empirical value for $\theta$, whereby $sin\theta = \frac{1}{2}; cos\theta = \frac{\sqrt{3}}{2}$, the predicted values of the experimental ratios now become
 
\begin{eqnarray}
  r^{s0f_0}_{sf_0} \equiv \frac{\Gamma(B \to f_0)}{\Gamma(B_s \to f_0)} \rightarrow \frac{1}{4+2\sqrt{3}}\nonumber \\
 r^{0f_0}_{0\sigma} \equiv \frac{\Gamma(B \to f_0)}{\Gamma(B \to \sigma)} \rightarrow \frac{1}{4}\nonumber \\
  r^{s\sigma}_{s f_0} \equiv \frac{\Gamma(B_s \to \sigma)}{\Gamma(B_s \to f_0)} = 0 \nonumber\\
  r^{sf_0}_{0\sigma} \equiv \frac{\Gamma(B_s \to f_0)}{\Gamma(B_d \to \sigma)} \rightarrow 2+\sqrt{3}
\label{eq:amps1}
\end{eqnarray}

The value $ r^{0f_0}_{0\sigma} \equiv \frac{\Gamma(B \to f_0)}{\Gamma(B \to \sigma)} \rightarrow \frac{1}{4}$ is consistent, at $1\sigma$ 
with the limit reported by LHCb.  We urge that LHCb refit their data along these lines.

There is an implication of our hypothesis for the production of $a_0$, which provides a further experimental test. We urge LHCb to study the ratio of $a_0/f_0$ 
production in each of $B_d$ and $B_s \to J/\psi + (a_0/f_0)$. We give the predicted values of these ratios for the flavour symmetric case ($\theta = \pi/4$) and show the change in this when $\theta \to \pi/6$.

\begin{eqnarray}
 \frac{\Gamma(B_d \to J/\psi a_0)}{\Gamma(B_d \to J/\psi f_0)} = 1 \rightarrow \frac{1}{3} \nonumber \\
\frac{\Gamma(B_s \to J/\psi a_0)}{\Gamma(B_s \to J/\psi f_0)} =0 \rightarrow (\frac{\sqrt{3}-1}{\sqrt{3}+1})^2 = 0.07
\label{eq:amps2}
\end{eqnarray}
The relative suppression of $a_0$ in $B_d \to J/\psi X$ and its appearance in $B_s \to J/\psi X$ are delicate measurements, but in principle feasible.

\section{summary of conclusions}

Our interpretation of the LHCb data on $B_{d,s} \to J/\psi 2\pi; 4\pi$ leads to the following qualitative conclusions.

1. The $f_1(1285)$ is consistent with the flavor mixture $f_1(1285) \sim 0.9n\bar{n} - 0.4 s\bar{s}$ \cite{ckf1}.

2. The data on $B_{d,s} \to J/\psi 4\pi$ show that $f_0(1370)$ and $f_0(1500)$ interfere, and that $s\bar{s}$ is more prominent in $f_0(1500)$ 
than in $f_0(1370)$ \cite{cqz}.

3. The data on $B_s \to J/\psi 2\pi$ show that there is a large $s\bar{s}$ component in $f_0(1710)$ and that this scalar interferes with the 
other scalar states and the S-wave background.

5. Thus we expect a prominent signal for $f_0(1710)$ in $B_{d,s} \to J/\psi K\bar{K}$. 
Evidence for a peak in $K\bar{K}$ spectrum is consistent with the parameters of the $f_0(1710)$ \cite{lhcbbskk}

\subsection{Further discussion}

The existing data on $B_{d,s} \to J/\psi + X$ are consistent with a canonical picture of scalar mesons, as deduced from historical data. 
We have used the historical picture to construct curves 
and compared them with LHCb data, where available. We have not made an attempt to perform a best fit to these data. 
It would therefore be interesting if LHCb now  made a fit to their acceptance corrected data taking into account our scenario. 
As the nature of the scalar mesons is so fundamental, not least in connection with the isolation of a scalar glueball degree of freedom in 
this mass region, the picture presented here merits serious examination.

We have given some further tests of our hypothesis, such as the production of $a_0$ in $B_d \to J/\psi X$. 
A further test of these ideas will come if neutrals can be detected and $\eta$s reconstructed. The spectrum for $B_{d,s} \to J/\psi \eta\eta$ would thus be valuable as an independent test of the flavour-glue mixing in the scalar mesons above 1 GeV.
A study of $B^{o,-} \to J/\psi \eta\pi$ is also relevant, for understanding the $a_0(980)$ production. 

In conclusion: the LHCb data appear to be consistent with the picture of scalar mesons below 1GeV being 
tetraquark states, and those above 1 GeV being a canonical nonet mixed with a scalar glueball.

\acknowledgments
We are grateful to Eric Swanson, Sheldon Stone and Greig Cowan for discussions during this analysis.

\end{document}